\documentclass[12pt,english]{article}
\usepackage[T1]{fontenc}
\usepackage[latin9]{inputenc}
\usepackage{datetime}
\usepackage{color}
\usepackage{amsthm}
\usepackage{amsmath}
\usepackage{amssymb}
\usepackage{cite}

  \theoremstyle{plain}
  
  \theoremstyle{plain}
  \newtheorem*{lem*}{Lemma}
\theoremstyle{plain}
\newtheorem{thm}{Theorem}

\usepackage{amsfonts}
\usepackage{amsthm}
\usepackage[labelfont=bf, font=small]{caption}

\makeatother

\usepackage{babel}

\begin{document}
\global\long\global\long\def\kb#1#2{|#1\rangle\langle#2|}
 \global\long\global\long\def\bk#1#2{\langle#1|#2\rangle}
 \global\long\global\long\def\ket#1{|#1\rangle}
 \global\long\global\long\def\bra#1{\langle#1|}
 \global\long\global\long\def\c#1{\mathbb{C}^{#1}}

\title{\textbf{\large The Limited Role of Mutually Unbiased Product Bases \\ in Dimension Six}}

\author{Daniel McNulty and Stefan Weigert\\
Department of Mathematics, University of York\\
York YO10 5DD, UK\\ \\
\small{\tt{dm575@york.ac.uk, stefan.weigert@york.ac.uk}}}

\date{13 March 2012}
\maketitle
\begin{abstract}
We show that a complete set of seven mutually unbiased bases in dimension six, if it exists, cannot contain more than one product basis. 
\end{abstract}


One way to express complementarity of quantum mechanical observables is to 
say that their eigenstates form a pair of mutually unbiased (MU) bases: if a system  resides in an eigenstate of one of these observables, the 
probability distribution to find the system in the eigenstates of the other observable is 
\emph{flat}. The state space of a $d$-level system 
accommodates maximally $(d+1)$ pairwise complementary observables known 
as a \emph{complete set} of MU bases which satisfy
\begin{equation} \label{eq: overlaps}
| \bk{j_a}{k_b}|^{2}
= \frac{1}{d}(1-\delta_{ab})+\delta_{jk}\delta_{ab},\quad
j,k=0\ldots d-1,\,a,b=0\ldots d\,,
\end{equation}
where the set $\{\ket{j_a}\}$ for fixed $a$ is one orthonormal basis of $\mathbb{C}^{d}$. If it exists, a complete set allows one to reconstruct the unknown quantum state of a $d$-level system with least statistical redundancy \cite{ivanovic81,wootters+89} and to set up secure methods of quantum key distribution \cite{cerf+02}, for example.

In prime-power dimensions $d=p^{k}$, with $k$ a positive integer, complete sets of MU bases have been constructed in a number of ways \cite{wootters+89,bandyo+02,klappenecker+04,durt04}. The uses and known properties of MU bases for discrete and continuous \cite{Wilkinson+09} variables have been reviewed  in \cite{Durt+10}.

For bipartite systems of composite dimension given by $d=pq$, with prime numbers $p<q$, say, complete sets of $(pq+1)$ MU bases have not been found, even for a quantum system with only six levels, i.e. $d=6$. In fact, all current evidence supports the conjecture \cite{zauner+99} that no more than \emph{three} MU bases exist in dimension six. Substantial numerical data \cite{butterley+07,brierley+08} seem to rule out the existence of more than three MU bases, while exact results drawn from both numerical calculations with rigorous error bounds \cite{Jaming+2009} and computer-algebraic methods \cite{grassl04,brierley+09} prove the impossibility to add more than one MU basis to specific given pairs. For the pair of MU bases corresponding to $\{ I,S_6\}$ it is not even possible to find a third MU basis \cite{brierley+10}; here $I$ is the identity matrix in $\mathbb{C}^{6}$ and $S_6$ is Tao's matrix \cite{tao04}.

The purpose of this contribution is to derive a rigorous result regarding the impossibility to extend certain pairs of MU bases in dimension six to complete sets. The special property of the MU bases we consider is that they only contain \emph{product} states $\ket{\psi,\Psi}\equiv \ket{\psi}\otimes \ket{\Psi}$ of the state space $\mathbb{C}^{6}$, with $\ket{\psi} \in \mathbb{C}^{2}$, and $\ket{\Psi} \in \mathbb{C}^{3}$. This approach complements studies of the entanglement structure of complete sets, mostly in prime power dimensions \cite{Romero+05,Wiesniak+11,Lawrence+11}. We will show that no \emph{pair} of MU product bases can figure in a complete set as stated by the following theorem.       

\begin{thm}\label{thm:noproductbases}
If a complete set of seven MU bases in dimension six exists, it contains at most one product basis.
\end{thm} 

This is, in fact, the \emph{strongest possible} bound on the number of MU product bases since one can always map one MU basis of a complete set to the standard basis. The proof will start from the exhaustive list of pairs of MU product bases of $\mathbb{C}^{6}$ constructed in \cite{mcnulty+11}. Not all of the listed pairs were given in the standard form which requires the first basis to be the computational basis \cite{weigert+10}. Thus, we will first bring the pairs of the list to standard form, using unitary equivalence transformations. We will find that the second MU product basis of each pair is mapped either to a member of the Fourier family of Hadamard matrices, discovered in \cite{haagerup97}, or to Tao's matrix \cite{tao04}. Using some of the results mentioned earlier, it is then straightforward to prove Theorem \ref{thm:noproductbases}.

To begin, we reproduce the set of pairs of MU product bases of a quantum system with six orthogonal states obtained in \cite{mcnulty+11}. They are expressed in terms of the complete sets of MU bases for $\mathbb{C}^{2}$ and $\mathbb{C}^{3}$, given by $\{\ket{j_z}\}$, $\{\ket{j_x}\}, \{\ket{j_y}\}$, and $\{\ket{J_z}\}, \{\ket{J_x}\}, \{\ket{J_y}\}, \{\ket{J_w}\}$, respectively. The bases consist of the eigenstates of the Heisenberg-Weyl operators $Z, X,Y$ $\equiv XZ$,  (and $W\equiv X^2Z)$ \cite{bandyo+02}, with clock and shift operators $Z$ and $X$ which satisfy $ZX=\omega XZ$, where $\omega = e^{2\pi i/d}$, $d=2,3$. 
\begin{thm}
\label{thm:All-pairs-d=00003D6} Any pair of MU product bases in the
space $\mathbb{C}^{2}\otimes\mathbb{C}^{3}$ is equivalent to a member
of the families \begin{align}
\mathcal{P}_{0} & =\{\ket{j_{z},J_{z}};\,\ket{j_{x},J_{x}}\}\,, \nonumber \\
\mathcal{P}_{1} &
=\{\ket{j_{z},J_{z}};\,\ket{0_{x},J_{x}},\ket{1_{x},\hat{R}_{\xi,\eta}J_{x}}\}\,,\nonumber\\
\mathcal{P}_{2} &
=\{\ket{0_{z},J_{z}},\ket{1_{z},J_{y}};\,\ket{0_{x},J_{x}},\ket{1_{x},J_{w}}\}\,,\nonumber \\
\mathcal{P}_{3} &
=\{\ket{0_{z},J_{z}},\ket{1_{z},\hat{S}_{\zeta,\chi}J_{z}};\,\ket{j_{x},0_{x}},\ket{\hat{r}_{\sigma}j_{x},1_{x}},\ket{\hat{r}_{\tau}j_{x},2_{x}}\}\,, \label{list}
\end{align}
 with $j=0,1$ and $J=0,1,2$. The unitary operator $\hat{R}_{\xi,\eta}$
is defined as
$\hat{R}_{\xi,\eta}=\kb{0_{z}}{0_{z}}+e^{i\xi}\kb{1_{z}}{1_{z}}+e^{i\eta}\kb{2_{z}}{2_{z}}\,,$
for $\eta,\xi\in[0,2\pi)$, and $\hat{S}_{\zeta,\chi}$ is defined
analogously with respect to the $x$-basis; the unitary operators
$\hat{r}_{\sigma}$ and $\hat{r}_{\tau}$ act on the basis
$\{\ket{j_{x}}\}\equiv\{\ket{\pm}\}$
according to $\hat{r}_{\sigma}\ket{j_{x}}=(\ket{0_{z}}\pm
e^{i\sigma}\ket{1_{z}})/\sqrt{2}$
for $\sigma\in(0,\pi)$, etc.
\end{thm}

The ranges of the six real parameters $\xi,\eta, \dots, \sigma, \tau$, are chosen in such a way that no MU product pair occurs more than once in the list (\ref{list}). For example, the operator $\hat{R}_{\xi,\eta}$ in $\mathcal{P}_{1}$ is required to be different from the identity in order not to reproduce the Heisenberg-Weyl pair $\mathcal{P}_{0}$. There are four sets of MU product pairs in dimension six but 
both $\mathcal{P}_{1}$ and $\mathcal{P}_{3}$ connect to $\mathcal{P}_{0}$, which is the only \emph{direct} product basis (cf. \cite{Wiesniak+11}) in the list while $\mathcal{P}_{2}$ is an \emph{isolated} pair.

It will be convenient to represent the MU product pairs of Theorem \ref{thm:All-pairs-d=00003D6} in terms of $(6 \times 6)$ unitary matrices. We associate the standard bases of $\mathbb{C}^{2}$ and $\mathbb{C}^{3}$ with $\{\ket{j_z}\}$ and $\{\ket{J_z}\}$, respectively. Then, $\{\ket{J_z}\}$ is represented by the identity $I_3$ and $\{\ket{J_x}\}$ by the Fourier matrix 
\begin{equation}
F_3=\frac{1}{\sqrt{3}}\left(\begin{array}{ccc}
1 & 1 & 1\\
1 & \omega & \omega^{2}\\
1 & \omega^{2} & \omega\end{array}\right)\,,
\label{xmatrix}\end{equation}
where $\omega=e^{2\pi i/3}$ is a third root of unity. The bases $\{\ket{J_y}\}$ and $\{\ket{J_w}\}$, both of which are MU to $\{\ket{J_z}\}$ and $\{\ket{J_x}\}$ and among themselves, are represented by the unitary matrices
\begin{equation}
H_{y}=\frac{1}{\sqrt{3}}\left(\begin{array}{ccc}
1 & 1 & 1\\
\omega & \omega^{2} & 1\\
\omega & 1 & \omega^{2}\end{array}\right)\,,\quad
H_{w}=\frac{1}{\sqrt{3}}\left(\begin{array}{ccc}
1 & 1 & 1\\
\omega^{2} & 1 & \omega\\
\omega^{2} & \omega & 1\end{array}\right)\,,\label{yandwmatrix}
\end{equation}
respectively. Thus, the MU product pairs given in (\ref{list}) can be represented by the following pairs of matrices, 
\begin{align}
\label{p0}
\mathcal{P}_0&= \{I;\,\widetilde{F}(0,0)\}\, ,\\
\label{p1}
\mathcal{P}_1&= \{I;\,\widetilde{F}^\text{T}(\xi,\eta)\}\, , \\
\label{p2}
\mathcal{P}_2&= \{\widetilde{I}(4\pi/3,4\pi/3);\,\widetilde{F}^{\text{T}}(4\pi/3,4\pi/3)\}\, ,\\
\label{p3}
\mathcal{P}_3&= \{\widetilde{I}(\zeta,\chi);\,\widetilde{F}(\sigma,\tau)\}\, .
\end{align}
Here, the unitary matrix $\widetilde{F}(\xi,\eta)$ is given by 
\begin{equation}\label{Ftilde}
\widetilde{F}(\xi,\eta)=\frac{1}{\sqrt{2}}\left(\begin{array}{cc}
F_3 & F_3\\
F_3D & -F_3D\end{array}\right),
\end{equation}
with $F_3$ from Eq. (\ref{xmatrix}) and a diagonal matrix $D=\text{diag}(1,e^{i\xi},e^{i\eta})$, a form occurring already in \cite{bengtsson+07}. The transpose of $\widetilde{F}$, present in $\mathcal{P}_1$ and $\mathcal{P}_2$, is denoted by $\widetilde{F}^\text{T}(\xi,\eta)$.
The family of non-standard bases $\widetilde{I}(\zeta,\chi)$ is given by 
\begin{equation}
\widetilde{I}(\zeta,\chi)=\left(\begin{array}{cc}
I_3 & 0\\
0 & S_{\zeta,\chi}\end{array}\right),
\quad \mbox{where}\quad
S_{\zeta,\chi}=\left(\begin{array}{cccccc}
a & c & b\\
b & a & c\\
c & b & a
\end{array}\right)\, , \label{matrixP_4}
\end{equation}
with
\begin{align}
a(\zeta,\chi)&=\frac{1}{3}(1+e^{i\zeta}+e^{i\chi})\, ,\\
b(\zeta,\chi)&=\frac{1}{3}(1+\omega^2e^{i\zeta}+\omega e^{i\chi})\, ,\\
c(\zeta,\chi)&=\frac{1}{3}(1+\omega e^{i\zeta}+\omega^2e^{i\chi})\, ,
\end{align}

$S_{\zeta,\chi}$ being diagonal in the eigenbasis of the operator $X$.

First, we show that the pair $\mathcal{P}_1 = \{I;\, \widetilde{F}^\text{T}(\xi,\eta) \}$ is equivalent to $\{I;\, \widetilde{F}(\xi,\eta) \}$. To see this we multiply the pair $\{I;\, \widetilde{F} \}$ with $\widetilde{F}^\dagger$, the adjoint of $\widetilde{F}$, from the left. The pair $\{I;\,\widetilde{F}\}$ becomes $\{\widetilde{F}^\dagger;\,I\}$, and taking the complex conjugate of the pair $\{\widetilde{F}^\dagger;\,I\}$ leaves us with $\{\widetilde{F}^{\text{T}};\,I\}$ which, after a swap, is indeed $\mathcal{P}_1$.  

Next, we show that the matrix $\widetilde{F}(\xi,\eta)$ is equivalent to the Fourier family of Hadamard matrices $F(\xi,\eta)$ as defined in \cite{tadej+06}. First we permute rows $2$ and $5$ of the matrix $\widetilde{F}(\xi,\eta)$, resulting in $\widetilde{F}^\prime(\xi,\eta)$, the columns of which are no longer product vectors. Then we reorder the columns of $\widetilde{F}^\prime$ such that columns $2,3,5$ and $6$ become columns $6,2,3$ and $5$, respectively, producing immediately the Fourier family $F(\xi,\eta)$. In a sense, we have \emph{derived} the Fourier family of Hadamard matrices through constructing MU product bases, thereby ``explaining'' why this set depends on two real parameters. Since the transformations just described do not affect the standard basis, we have shown the equivalence of    
$\mathcal{P}_1$ with the pair $\{I;\, F(\xi,\eta) \}$.

Now we will show  that the pair $\mathcal{P}_3\equiv\{\widetilde{I}(\zeta,\chi);\, \widetilde{F}(\sigma,\tau)\}$ is also equivalent to $\mathcal{P}_1$. To see this, we transform the first basis $\widetilde{I}(\zeta,\chi)$ into the identity by multiplying it from the left with its inverse,
\begin{equation}
\left(\begin{array}{cc}
I_3 & 0\\
0 & S_{\zeta,\chi}^{\dagger}\end{array}\right),\label{eq:unitaryd=6}\end{equation}
where $S_{\zeta,\chi}^{\dagger}$ is the adjoint of $S_{\zeta,\chi}$, defined in $(\ref{matrixP_4})$, simultaneously mapping the matrix $\widetilde{F}(\sigma,\tau)$ (see Eq. (\ref{Ftilde}) to
\begin{equation}
\frac{1}{\sqrt{2}}\left(\begin{array}{cc}
F_3 & F_3\\
S_{\zeta,\chi}^{\dagger}F_3D & -S_{\zeta,\chi}^{\dagger}F_3D\end{array}\right) \, .
\end{equation}
Since $S_{\zeta,\chi}$ is diagonal in the $X$ basis, $S_{\zeta,\chi}^\dagger$ simply multiplies the columns of each matrix $F_3$ by phase factors. Writing $\sigma^\prime=\sigma-\zeta$, we obtain the desired equivalence
\begin{equation} \label{sim1and3}
\mathcal{P}_3 
\sim \{ I; \,\widetilde{F}(\sigma-\zeta,\tau-\chi)\} 
= \{ I; \,\widetilde{F}(\sigma^\prime,\tau^\prime)\} 
\sim \mathcal{P}_1 \, .
\end{equation}

Finally, we show that $\mathcal{P}_2$ is equivalent to the pair 
$\{I;\,S_6\}$. Expressing the pair as 
\begin{equation}
\mathcal{P}_2 = \left\{\left( \begin{array}{cc}
I_3 & 0 \\
0 & -iH_y \end{array} \right);\,
\frac{1}{\sqrt{2}}\left( \begin{array}{cc}
F_3 & H_w\\
F_3 & -H_w \end{array} \right)\right\},
\end{equation}
with matrices $H_y$ and $H_w$ defined in Eq. $(\ref{yandwmatrix})$, suggests to map the first matrix to the identity by multiplying it with 
\begin{equation}
\left(\begin{array}{cc}
I_3 & 0\\
0 & iH^{\dagger}_y\end{array}\right)\label{eq:unitaryd=6number2}
\end{equation}
from the left. The second matrix of $\mathcal{P}_2$ turns into
\begin{equation}
\widetilde{S}_6=\frac{1}{\sqrt{2}}\left(\begin{array}{cc}
F_3 & H_w\\
iH^{\dagger}_yF_3 & -iH^{\dagger}_yH_w\end{array}\right),
\end{equation}
with
\begin{equation}
iH^{\dagger}_yF_3=\frac{1}{\sqrt{3}}\left(\begin{array}{ccc}
1 & \omega & \omega\\
\omega & 1 & \omega\\
\omega & \omega & 1\end{array}\right)\quad\mbox{and}\quad
iH^{\dagger}_yH_w=-\frac{1}{\sqrt{3}}\left(\begin{array}{ccc}
1 & \omega^2 & \omega^2\\
\omega^2 & 1 & \omega^2\\
\omega^2 & \omega^2 & 1\end{array}\right).
\end{equation}

To transform $\widetilde{S}_6$ into the Hadamard matrix $S_6$ we perform a number of  simple operations. First we swap the second row of $\widetilde{S}_6$ with its third row as well as its fourth and fifth rows. Then we permute columns two with six, three with five, and four with five, followed by a multiplication of rows four and six by $\omega^2$. These equivalence transformations indeed result in the matrix $S_6$ while their action on the identity is easily undone by column operations, thus establishing the equivalence relation 
\begin{equation}
\mathcal{P}_2\sim\{I;\,S_6\} \, ,
\end{equation}
which concludes the simplification of the list of MU product pairs. As with the Fourier family, we have ``derived'' Tao's matrix $S_6$ from a pair of MU product bases.

To summarize, the standard form of the set of MU product pairs listed in Eqs. $(\ref{p0})$-$(\ref{p3})$ reduces to
\begin{eqnarray} 
\mathcal{P}_0&\sim&\{I;\,F(0,0)\} \, , \nonumber \\
\mathcal{P}_1\sim\mathcal{P}_3&\sim&\{I;\,F(\xi,\eta)\} \, , \nonumber \\
\mathcal{P}_2&\sim&\{I;\,S_6\} \, , \label{reducedlist} 
\end{eqnarray}
with $\mathcal{P}_1$ and $\mathcal{P}_3$ equivalent to a two-parameter family and $\mathcal{P}_2$ being an isolated pair. 

It is now straightforward to complete the proof of Theorem \ref{thm:noproductbases}. Using computer-algebraic methods, it has been shown that the standard basis together with the isolated Hadamard $S_6$ \emph{cannot} be extended to a triple of MU bases: there are 90 vectors MU to $\{I;\,S_6\}$ \cite{brierley+09} but no two of them are orthogonal \cite{brierley+10}. Thus, $\mathcal{P}_2$ cannot figure in a complete set of seven MU bases. Combining numerical calculations with rigorous error bounds \cite{Jaming+2009}, all pairs of MU bases involving members of the Fourier family\footnote{In terms of our conventions, the result \cite{Jaming+2009} applies to the \emph{transposed} Fourier family, i.e. directly to the pair $\mathcal{P}_1$ in Eq. (\ref{p1}).} have been shown rigorously not to extend to quadruples of MU bases. These two results cover all cases given in (\ref{reducedlist}), hence all MU product pairs of the list (\ref{list}). It follows that \emph{no complete set of seven MU bases in $d=6$ contains a pair of MU product bases}, i.e. Theorem \ref{thm:noproductbases}.


We set out in \cite{mcnulty+11} with the modest goal to construct all MU product basis in dimension six. Using the resulting exhaustive list of MU product pairs, we have now been able to conclude that six of the seven MU bases required for a complete set in $\mathbb{C}^{6}$ must contain entangled states - if such a set exists. To our knowledge, this is the strongest rigorous result concerning the structure of MU bases for $d=6$. It considerably generalises the result that no pair of MU bases associated with the Heisenberg-Weyl operators of $\mathbb{C}^{6}$ can give rise to a complete set \cite{grassl04}, at the same time providing an independent proof thereof. It is also stronger than a result given in \cite{Wiesniak+11}, where the fixed entanglement content of a complete set in $d=6$ has been used to show that no more than \emph{three} of the seven hypothetical MU bases can be product bases. In addition, the current approach sheds some light on the particular character of the Fourier family of Hadamard matrices and Tao's matrix, since these - and only these - matrices emerge naturally upon constructing all pairs of MU product bases in dimension six.

\subsubsection*{Acknowledgements}
We thank M. Matolcsi to confirm that the result \cite{Jaming+2009} applies to the Fourier family \emph{and} the transposed Fourier family.  This work has been supported by EPSRC.

\end{document}